\documentclass[a4paper,12pt]{article}
\usepackage[dvipdfmx]{graphicx}
\usepackage{bmpsize}
\usepackage{natbib}
\usepackage{setspace}
\usepackage{multirow}
\usepackage{lmodern}
\usepackage{amsmath, amssymb}
\usepackage{bm}
\usepackage{pdflscape}
\usepackage{color}
\usepackage{booktabs}
\usepackage{afterpage}
\usepackage{textcomp}
\usepackage{float}
\usepackage{enumerate}

\makeatletter
\newcommand{\figcaption}[1]{\def\@captype{figure}\caption{#1}}
\newcommand{\tblcaption}[1]{\def\@captype{table}\caption{#1}}
\makeatother
\begin{document}
\title{Scalable Inference for Space-Time Gaussian Cox Processes}
\author{Shinichiro Shirota \thanks{Department of Biostatistics, UCLA., US. E-mail:shinichiro.shirota@gmail.com} and Sudipto Banerjee \thanks{Department of Biostatistics, UCLA., US. E-mail:sudipto@ucla.edu}}
\maketitle
\begin{abstract}
The log-Gaussian Cox process is a flexible and popular class of point pattern models for capturing spatial and space-time dependence for point patterns.
Model fitting requires approximation of stochastic integrals which is implemented through discretization over the domain of interest. With fine scale discretization, inference based on Markov chain Monte Carlo is computationally burdensome because of the cost of matrix decompositions and storage, such as the Cholesky, for high dimensional covariance matrices associated with latent Gaussian variables. This article addresses these computational bottlenecks by combining two recent developments: (i) a data augmentation strategy that has been proposed for space-time Gaussian Cox processes that is based on exact Bayesian inference and does not require fine grid approximations for infinite dimensional integrals, and (ii) a recently developed family of sparsity-inducing Gaussian processes, called nearest-neighbor Gaussian processes (NNGP), to avoid expensive matrix computations. Our inference is delivered within the fully model-based Bayesian paradigm and does not sacrifice the richness of traditional log-Gaussian Cox processes. We apply our method to crime event data in San Francisco and investigate the recovery of the intensity surface. 

{\it keywords:} Gaussian Cox processes, Gaussian processes, nearest neighbor Gaussian processes (NNGP), Poisson thinning, space-time point pattern
\end{abstract}
\section{Introduction}
The modeling and analysis for space and space-time point pattern data continue to be of interest in diverse settings including, but not limited to, point patterns of locations of tree species \citep[see, e.g.,][]{Burslemetal(01), Wiegandetal(09), Illianetal(08)}, locations of disease occurrences \citep{Liangetal(09),RuizMorenoetal(10),Diggleetal(13)}, locations of earthquakes \citep[][]{Ogata(99),MarsanLengline(08)} and locations of crime events \citep[][]{ChaineyRatcliffe(05), GrubesicMack(08), ShirotaGelfand(17a)}. In addition, the points may be observed over time \citep{GrubesicMack(08),Diggleetal(13)}. 
General theory on point processes can be found in texts such as \cite{DaleyVereJones(03)} and \cite{DaleyVereJones(08)}, while spatial point patterns have been specifically discussed in \cite{Lantuejourl(02)}, \cite{Illianetal(08)}, \cite{Gelfandetal(10)}, \cite{Diggle(13)}, and \cite{Baddeleyetal(15)}. These also contain dependent time series modeling of spatial point patterns.  The current literature has tended to focus primarily on nonhomogeneous Poisson processes (NHPP) or, more generally, log Gaussian Cox processes (LGCP) \citep[see, e.g.,][and references therein]{MollerWaagepetersen(04)}. 
The intensity surface of a Cox process is treated as a realization of a stochastic process, which captures stochastic spatial and space-time dependence.
Given the intensity surface, Cox processes are Poisson processes. The LGCP was originally proposed by \cite{Molleretal(98)} and extended to the space-time case by \cite{BrixDiggle(01)}. As the name suggests, the intensity function of the LGCP is driven by the exponential of a Gaussian processes (GP).

Fitting LGCP models is challenging because the likelihood of the LGCP involves integrating the intensity function over the domain of interest. The integral is stochastic and is analytically intractable, so some approximations are required. One customarily grids the study region (creating a set of so called representative points) by $K$ tiles and approximate this integral with a Riemann sum \citep[][]{Molleretal(98),MollerWaagepetersen(04)}. Typically, a large number of tiles, i.e., large values of $K$, are required for accurate inference. Bayesian model fitting provides richer and more flexible inference and is typically achieved using Markov chain Monte Carlo (MCMC) methods. However, these are computationally more demanding because they require repeated approximations for a very large number of MCMC iterations to satisfy adequate convergence. Moreover, a standard MCMC scheme needs repeated conditional sampling of high dimensional latent Gaussian variables. 
However, computing with high dimensional GPs remains demanding. Typically a determinant and a quadratic form involving the inverse of the space-time covariance matrix is required. The Cholesky decomposition of the covariance matrix is a customary choice that delivers the determinant and inverse. For a $K$-dimensional covariance matrix, these calculations require floating point operations (flops) in the order of $\mathcal{O}(K^{3})$ and $\mathcal{O}(K^{2})$ memory for storage.

An alternative approach is to employ the sigmoidal Gaussian Cox processes (SGCP) proposed by \cite{AdamsMurrayMacKay(09)}. This approach utilizes the thinning property for NHPP (\cite{LewisShedler(79)}) to avoid any grid approximations. It obtains exact inference by introducing and sampling a GP on latent thinned points in addition to observed points. Although this does not require evaluating the intractable stochastic integral, sampling a GP on observed and latent thinned points is necessary within each MCMC iteration. When the number of observed points is large or the intensity surface is highly peaked in small areas, implementation of this approach can be computationally infeasible. 

Recently, exact space-time Gaussian Cox processes (we call exGCP in this paper) were proposed by \cite{GoncalvesGamerman(18)}. The idea of this approach is similar to SGCP, but they consider the Gaussian distribution function instead of the sigmoidal function. This approach also avoids high dimensional tiled surfaces, but still requires matrix factorizations that can become expensive with a large number of points. The number of points considered by \cite{GoncalvesGamerman(18)} was not large. Scaling up the algorithm is one of the promising directions for applying this method to large point pattern datasets. There is, by now, a burgeoning literature on efficiently handling GPs for large spatial datasets. A comprehensive review is beyond the scope of the current article; see recent review articles by \cite{SunLiGenton(12)} and \cite{Banerjee(17)}. A recent ``contest'' paper by \cite{HeatonContest(17)} shows many methods, including the one we adopt here, to be very competitive and delivering effectively indistinguishable inference on the spatial process.

One approach that is receiving much traction in high-dimensional spatial statistics is based upon \cite{Vecchia(88)}, who proposed a computationally efficient likelihood approximation based upon what could be characterized as a directed acyclic graph, or DAG, decomposition of the joint multivariate Gaussian density exploiting a much smaller set of conditional variables determined from nearest neighbors. This idea is now commonly used in graphical Gaussian models to introduce sparsity in the precision matrix. \cite{Dattaetal(16a)} extended this likelihood approximation to a sparsity-inducing Gaussian process, calling it a Nearest-Neighbor Gaussian Process (NNGP), enabling spatial prediction and interpolation at arbitrary locations. The resulting sparse precision matrix for the realizations of this process is available in closed form up to the process parameters and allows for very fast computations. The NNGP's role as an efficient Bayesian model relies upon the well-established accuracy and computational scalability of Vecchia's approach, which has also been demonstrated by several authors including \cite{Steinetal(04)} and more recently by \cite{Guinness(18)}. The potential for scalability is massive as the computational complexity is $O(KM^3)$, i.e., linear in the number of points $K$, which is usually large, and cubic in $M$ which is the fixed number of neighbors and is usually fixed at a small number. For example, \cite{Finleyetal(17)} present different classes of NNGP specifications and show that $M=10$ or $20$ is sufficient for approximating GP realizations over millions of locations.


In the current manuscript we propose scalable inference for large space-time point patterns by incorporating NNGP specifications into the exGCP framework of \cite{GoncalvesGamerman(18)}. Replacing the GP with an NNGP accrues computational benefits while ensuring valid probability models. We investigate recovering the GP and intensity surface through simulation studies and also apply our model to analyze crime event data in San Francisco (SF). The format of the paper is as follows. Section~2 reviews some Bayesian inference approaches for LGCPs. Section~3 introduces the space-time exGCP by \cite{GoncalvesGamerman(18)} and NNGP by \cite{Dattaetal(16a)}. In Section~4, we discuss Bayesian inference and NNGP implementation for the model and their computational complexity. Section~5 provides simulation studies to demonstrate the intensity recovery. In Section~6, the model is implemented for the crime event data in San Francisco. Finally, Section~7 offers some discussion and concluding remarks.

\section{Cox Processes driven by Gaussian processes: A brief review}
Let $\mathcal{S}=\{\bm{s}_{1},\ldots, \bm{s}_{n}\}$ be an observed point pattern on $\mathcal{D}\subset \mathbb{R}^{d}$, where $\mathcal{D}$ is a bounded study region. A simple point process model is the nonhomogeneous Poisson process (NHPP), with likelihood 
\begin{align}
\mathcal{L}(\mathcal{S}|\lambda(\cdot)) &= \exp\biggl(|\mathcal{D}|-\int_{\mathcal{D}}\lambda(\bm{u})d\bm{u}\biggl)\prod_{i=1}^{n}\lambda(\bm{s}_{i}), \quad \log \lambda(\bm{s})=\bm{X}(\bm{s})\bm{\beta}\;,
\end{align}
where $\lambda(\cdot)$ is a deterministic intensity surface and $\bm{X}(\cdot)$ is a covariate surface. This likelihood is analytically intractable because it involves $\int_{\mathcal{D}}\lambda(\bm{u})d\bm{u}$ which, in general, cannot be calculated explicitly. For further details on the NHPP, we refer to \cite{Illianetal(08)} and references therein. 

Cox processes are defined as point processes with a stochastic intensity surface. Thus, $\lambda(\cdot)$ is driven by some stochastic processes. 
The most popular specification is known as the log Gaussian Cox process (LGCP) proposed by \cite{Molleretal(98)}, which assumes that the logarithm of intensity surface $\lambda(\cdot)$ is driven by a GP. Therefore, 
\begin{align}
\log \lambda(\bm{s})=\bm{X}(\bm{s})\bm{\beta}+z(\bm{s}), \quad \bm{z}\sim \mathcal{N}(\bm{0}, \mathbf{C}_{\bm{\theta}}(\mathcal{S}, \mathcal{S}'))\;,
\end{align}
where $\bm{z}$ is an $n$-dimensional Gaussian random variable with mean $\bm{0}$ and covariance matrix $\mathbf{C}_{\bm{\theta}}(\mathcal{S}, \mathcal{S}')=[C_{\bm{\theta}}(\bm{s}_{i}, \bm{s}_{j})]_{i,j=1,\ldots,n}$. Generally, for Bayesian inference, we need to approximate $\int_{\mathcal{D}}\lambda(\bm{s})d\bm{s}$ to compute the likelihood. Specifically, we seek $\int_{\mathcal{D}}\lambda(\bm{s})d\bm{s}\approx \sum_{k=1}^{K}\lambda(\bm{s}_{k}^{*})\Delta_{k}$ where $\bm{s}_{k}^{*}$ and $\Delta_{k}$ are representative points and the area of grid $k$, respectively. This approximation results in the following likelihood representation,
\begin{align}
\mathcal{L}(\mathcal{S}|\lambda(\cdot))&\propto \exp\biggl(-\sum_{k=1}^{K}\lambda(\bm{s}_{k}^{*})\Delta_{k} \biggl)\prod_{k=1}^{K}\lambda(\bm{s}_{k}^{*})^{n_{k}}, 
\end{align}
where $n_{k}$ is the number of points in grid $k$, i.e., $\sum_{k=1}^{K}n_{k}=n$. Large values of $K$ are usually required for accurate Bayesian inference. This still creates a problem because $K$ determines the size of the covariance matrix whose inverse and determinant will be required in Bayesian computations. Without any exploitable structure, the computational cost is $\mathcal{O}(K^3)$. Some GP sampling methods have been investigated in the context of NHPPs. 
These include elliptical slice sampling \citep[][]{MurrayAdamsMacKay(10),LeiningerGelfand(17)},  Metropolis adjusted Langevin algorithm \citep[MALA, e.g., ][]{Besag(94), Molleretal(98), RobertsTweedie(96b)} and Riemann manifold MCMC \citep{GirolamiCalderhead(11)}, but computational costs still hover around $\mathcal{O}(K^3)$ without further assumptions on the GP. 
To complicate matters, the results can be sensitive to the grid approximation \citep[][]{Simpsonetal(16b)} and it is difficult to quantify the bias resulting from the grid. Furthermore, the number of grids is often unknown and can be specific to the application at hand. 

Integrated Nested Laplace Approximation \citep[INLA,][]{Rueetal(09)} is a computationally efficient approximate Bayesian inference for latent GP models. 
This approach approximates a precision matrix of a GP by Gaussian Markov random fields \citep[GMRF, see, e.g., ][]{RueHeld(05)}, whose computational cost is $\mathcal{O}(K^{3/2})$ and $\mathcal{O}(K\log(K))$ dynamic memory storage. A software package also has been developed \citep{LindgrenRue(15)}.
\cite{Illianetal(12)} investigate the INLA framework for the LGCP context, especially to large point patterns and a point pattern with multiple marks. 
\cite{TaylorDiggle(14)} compare INLA approach with MALA, demonstrate predictive outperformance of MALA to INLA.  
\cite{Brown(15)} make the interface to functions from the \texttt{INLA} package for spatial LGCP inference. \cite{Tayloretal(15)} provide a software package for Bayesian inference (MALA and INLA) of spatiotemporal and multivariate LGCP. 

Within a classical inferential paradigm, a minimum contrast estimator \citep[MCE, see, e.g., ][and references therein]{Illianetal(08)} has been investigated and implemented for the LGCP \citep[][]{Molleretal(98)}. 
This estimator is obtained by minimizing the distance of some parametric functional summary statistics, e.g., $K$-function and $L$-function, to their empirical estimators with respect to parameter values. 
These are easily implementable as long as a closed form of functional summary statistics is available, but are implemented for second order moments that do not generally characterize the distribution completely. 
The distribution of the LGCP is completely determined by its first and second order properties, so MCE is a practically useful approach for estimating parameters \citep[][]{Molleretal(98)}.

An alternative Bayesian approach is to employ the sigmoidal Gaussian Cox processes (SGCP) proposed by \cite{AdamsMurrayMacKay(09)}.
This approach utilizes the thinning property for NHPP (\cite{LewisShedler(79)}) to avoid any grid approximations. We achieve exact inference by introducing and sampling GPs on latent thinned points in addition to observed points. The sigmoidal Gaussian Cox processes (SGCP) by \cite{AdamsMurrayMacKay(09)} specifies the intensity as $\lambda(\cdot)=\lambda^{*}\varphi[z(\cdot)]$, where $\lambda^{*}$ is an upper bound on the intensity surface over the study region and $\varphi[\cdot]$ is the logistic function, $\varphi[z]=(1+\exp(-z))^{-1}$. These authors introduce latent points, $\mathcal{U}=\{\bm{u}_{1},\ldots,\bm{u}_{m}\}$, and consider $\mathcal{S}_{aug}=\{\mathcal{S}, \mathcal{U} \}$ as a realization from a homogeneous Poisson point process over $\mathcal{D}$ with intensity $\lambda^{*}|\mathcal{D}|$, where $|\mathcal{D}|$ is the area of $\mathcal{D}$.
Then, the joint density of $\{\mathcal{S}_{aug}, m, \bm{z}(\mathcal{S}_{aug}), \lambda^{*}\}$ is
\begin{align}
\mathcal{L}(\mathcal{S}_{aug}, m, \bm{z}(\mathcal{S}_{aug}), \lambda^{*}|\bm{\theta})&\propto \frac{(\lambda^{*})^{n+m}}{(n+m)!}\exp\{-\lambda^{*}|\mathcal{D}|\}\prod_{i=1}^{n}\varphi[z(\bm{s}_{i})]
\prod_{j=1}^{m}\varphi[-z(\bm{u}_{j})] \nonumber \\
&\times \mathcal{N}(\bm{z}(\mathcal{S}_{aug})|\bm{0},\mathbf{C}_{\bm{\theta}}(\mathcal{S}_{aug},\mathcal{S}_{aug}^{'}))
\end{align}
where $\bm{z}(\mathcal{S}_{aug})$ is an $(n+m)\times 1$ Gaussian random vector on $\mathcal{S}_{aug}$ and $\mathbf{C}_{\bm{\theta}}(\mathcal{S}_{aug}, \mathcal{S}_{aug}^{'})$ is the $(n+m)\times (n+m)$ covariance matrix.

This specification suggests that the $n+m$ points are uniformly generated by a homogeneous Poisson process with the intensity $\lambda^{*}$ over $\mathcal{D}$. Then, $\mathcal{S}$ is considered as a set of observed points and $\mathcal{U}$ is a set of unobserved thinned events with probability $\varphi[\cdot]$ through the thinning property for NHPP (\cite{LewisShedler(79)} and \cite{Ogata(81)}) with the intensity surface $\lambda^{*}\varphi[z(\cdot)]$. In addition to $\bm{\theta}$, ($m$, $\mathcal{U}$, $\bm{z}(\mathcal{S}_{aug})$, $\lambda^{*}$) are updated using MCMC. Although this approach does not require computing the stochastic integral, the sampling of an $n+m$ dimensional vector from the GP is necessary within each MCMC iteration. When $n$ is large or the intensity surface is highly peaked on subregions (more events will be retained under the thinning), this algorithm can become computationally unfeasible.

A promising recent development is by \cite{GoncalvesGamerman(18)}, who propose exGCP with intensity $\lambda(\cdot)=\lambda^{*}\Phi[z(\cdot)]$, where $\Phi$ is the cumulative distribution function of the standard Gaussian distribution. They consider a data augmentation strategy similar to \cite{AdamsMurrayMacKay(09)}, i.e., introducing latent thinned events $\mathcal{U}$ to avoid evaluation of $\int_{\mathcal{D}}\lambda(\bm{u})d\bm{u}$.  They also propose an exact Gibbs sampling algorithm for $m$, $\mathcal{U}$, $\bm{z}(\mathcal{S}_{aug})$, $\lambda^{*}$ and demonstrate that the algorithm is highly efficient for applications with a relatively small numbers of points. Although the approach is exact and highly efficient, the computational cost is still $\mathcal{O}((n+m)^3)$, which, once again, precludes modeling point patterns with very large number of points. 

To summarize, the LGCP is versatile and rich in its inferential capabilities, although the tiled surface approximations generate biases when $K$ is not large enough.  \cite{GoncalvesGamerman(18)} offers exact inference for these types of models using Gibbs sampling for $\bm{z}$ with data augmentation. The computational bottleneck of \cite{GoncalvesGamerman(18)} stems from required matrix factorizations for GP models. We consider a sparsity-inducing GP for scaling up Bayesian inference for the model.
\section{Scalable Space-Time Gaussian Cox Processes}
We now turn to scalable inference for the exGCP model in space-time contexts. 
\subsection{Space-time Gaussian Cox processes} 
We follow the specification by \cite{GoncalvesGamerman(18)} for the space-time exGCP. They assume the case of continuous space and discrete time, which is often appropriate for observed environmental processes (see, e.g., \cite{BanerjeeCarlinGelfand(14)}). Let $\mathcal{T}=\{1,2,\ldots,T\}$ be a set of time indices, $\mathcal{S}_{t}=\{\bm{s}_{t,1},\ldots,\bm{s}_{t,n_{t}}\}$ be an observed point pattern at time $t$, $n_{t}$ be the number of points in $\mathcal{S}_{t}$ and let us define $\mathcal{S}=\{\mathcal{S}_{1},\ldots, \mathcal{S}_{T}\}$.
Extensions of a GP to cope with space and discrete time were considered by \cite{Gelfandetal(05c)}; $\bm{z}$ follows a dynamic GP in discrete time if it can be described by a difference equation
\begin{align*}
\bm{z}_{t+1}&=\mathbf{G}\bm{z}_{t}+\bm{\eta}_{t}, \quad \bm{\eta}_{t}\sim \mathcal{N}(\bm{0}, \mathbf{C}_{\bm{\theta}_{t}}), \quad \text{for} \quad t=2,\ldots, T \\ 
\bm{z}_{1}&=\bm{\eta}_{1}, \quad \bm{\eta}_{1}\sim \mathcal{N}(\bm{0}, \mathbf{C}_{\bm{\theta}_{1}})
\end{align*}
where $\bm{\eta}_{t}$ is considered as an independently distributed Gaussian noise vector with covariance $\mathbf{C}_{\bm{\theta}_{t}}$ for $t=2,\ldots, T$. Similar processes were proposed in continuous time by \cite{BrixDiggle(01)}. Several options are available for the temporal transition matrix $\mathbf{G}$, e.g., autoregressive coefficient and identity matrix. 

Let $\bm{W}_{t}(\bm{s})=(1, X_{t,1}(\bm{s}), \ldots, X_{t,p}(\bm{s}))$ and $\bm{\beta}_{t}(\bm{s})=(z_{t}(\bm{s}), \beta_{t,1}(\bm{s}), \ldots,\beta_{t,p}(\bm{s}))^{\top}$ where $X_{t,j}(\bm{s})$ is the $j$th component of $\bm{X}_{t}(\bm{s})$ and $\beta_{t,j}(\bm{s})$ is the corresponding coefficient, the model is defined as
\begin{align}
\begin{split}
\mathcal{L}(\mathcal{S}|\lambda(\cdot))&\propto \exp\biggl(-\sum_{t=1}^{T}\int_{\mathcal{D}}\lambda_{t}(\bm{u})d\bm{u} \biggl)\prod_{t=1}^{T}\prod_{i=1}^{n_{t}}\lambda_{t}(\bm{s}_{t,i}) \\
\lambda_{t}(\bm{s})&=\lambda_{t}^{*}\Phi[f(\bm{W}_{t}(\bm{s}), \bm{\beta}_{t}(\bm{s}))], \quad f(\bm{W}_{t}(\bm{s}), \bm{\beta}_{t}(\bm{s}))=\bm{W}_{t}(\bm{s})\bm{\beta}_{t}(\bm{s}), \\ 
\bm{z}_{t}&=\mathbf{G}\bm{z}_{t-1}+\bm{\eta}_{t}, \quad \bm{\eta}_{t} \sim \mathcal{N}(\bm{0}, \mathbf{C}_{\bm{\theta}_{t}}), \quad \text{for} \quad t=2,\ldots, T \\ 
\bm{z}_{1}&=\bm{\eta}_{1}, \quad  \bm{\eta}_{1}\sim \mathcal{N}(\bm{0}, \mathbf{C}_{\bm{\theta}_{1}})
\end{split}
\end{align}
where $\Phi[\cdot]$ is the cumulative distribution function of the standard Gaussian distribution. 
Without loss of generality, we set $\mathbf{G}=\mathbf{I}$ in the below discussion.
We also assume $\bm{\theta}_{2}=\cdots=\bm{\theta}_{T}=\bm{\theta}$ and set different values for $\bm{\theta}_{1}$ so that covariance function $C_{\bm{\theta}_{1}}$ should have larger variance and stronger spatial dependence than $C_{\bm{\theta}}$. 
$\bm{\eta}_{t}$ for $t=2,\ldots, T$ capture the difference between $\bm{z}_{t}$ and $\mathbf{G}\bm{z}_{t-1}$, which can be considered weakly spatially correlated noise and be expected to have smaller spatial correlation and variance than $\bm{\eta}_{1}$. 

\subsection{Nearest neighbor Gaussian processes} 
In general, scalable GP models are constructed based upon low-rank approaches, sparsity-inducing approaches or some combination thereof. Low-rank models attempt to construct spatial GP on a lower-dimensional subspace using basis function representations \citep[see, ][and references therein]{Wikle(11)}. 
The computational cost for model fitting decreases from $\mathcal{O}(n^3)$ to $\mathcal{O}(nr^2)$ flops, where $r$ is the dimension of the lower-dimensional subspace or, equivalently, the number of basis functions. 
However, when $n$ is large, empirical investigations indicate that $r$ must be large to adequately approximate the original process impairing scalability to large datasets. 

An alternative is to develop full rank models that exploit sparsity. Covariance tapering \citep{Furreretal(06),Kaufmanetal(08)} introduces sparsity in the spatial covariance matrix $\mathbf{C}_{\bm{\theta}}$ using compactly supported covariance functions. 
This is effective for parameter estimation and interpolation of the response, but it has not been explored in depth for more general inference on residual or latent processes, as is required in our current setting with exGCPs. More recently, \cite{Dattaetal(16a)} proposed the NNGP approach, whose finite-dimensional realizations have sparse precision matrices available in closed form. 
The idea extends the principle of likelihood approximations outlined in \cite{Vecchia(88)} using directed acyclic graphs or Bayesian networks (terms not used by Vecchia) with parent sets comprising smaller sets of locations.  We review this briefly below.  

Sparsity itself has been effectively exploited \citep[][]{Vecchia(88), Steinetal(04), GramacyApley(15)} for approximating expensive likelihoods. A fully process-based modeling and inferential framework was proposed by \cite{Dattaetal(16a)}. Sparsity is typically introduced in the precision matrix $\mathbf{C}_{\bm{\theta}}^{-1}$ to approximate GP likelihoods \citep[see, e.g.,][]{RueHeld(05)} using, for example, the INLA algorithms \citep{Rueetal(09)}. However, this approach may produce biases, albeit often small, due to approximations and unlike low rank processes, these do not, necessarily, extend inference to new random variables at arbitrary locations without adding to the computational burden.

NNGP expresses the joint density of $\bm{z}$ as the product of approximated conditional densities by projecting on {\it neighbors} instead of the full set of locations, i.e., 
\begin{align}
\pi(\bm{z}(\mathcal{S}))&=\pi(z(\bm{s}_{1}))\pi(z(\bm{s}_{2})|z(\bm{s}_{1}))\cdots \pi(z(\bm{s}_{i})|\bm{z}_{<i})\cdots \pi(z(\bm{s}_{n})|\bm{z}_{<n}) \nonumber \\
&\approx \pi(z(\bm{s}_{1}))\pi(z(\bm{s}_{2})|z(\bm{s}_{1}))\cdots \pi(z(\bm{s}_{i})|\bm{z}_{N_{i}})\cdots \pi(z(\bm{s}_{n})|\bm{z}_{N_{n}})=\tilde{\pi}(\bm{z}(\mathcal{S}))
\end{align}
where $\bm{z}_{<i}=\{z(\bm{s}_{1}),\ldots, z(\bm{s}_{i-1}) \}$ and $N_{i}$ is the set of indices of neighbors of $\bm{s}_{i}$, $\bm{z}_{N_{i}}\subseteq \bm{z}_{<i}$ (see, e.g., \cite{Vecchia(88)}, \cite{Steinetal(04)}, \cite{Gramacyetal(14)} and \cite{GramacyApley(15)}). 
$\tilde{\pi}(\bm{z}(\mathcal{S}))$ is a proper multivariate joint density (\cite{Dattaetal(16a)}). 
As for neighbor selections, choosing $N_{i}$ to be any subset of $\{\bm{s}_{1},\ldots,\bm{s}_{i-1}\}$ ensures a valid probability density. For example, \cite{Vecchia(88)} specified $N_{i}$ to be the $M$ nearest neighbors of $\bm{s}_{i}$ among $\{\bm{s}_{1},\ldots,\bm{s}_{i-1}\}$ with respect to Euclidean distance. 
Sampling from $\tilde{\pi}(\bm{z}(\mathcal{S}))$ is sequentially implemented for $i=1,\ldots, n$ by drawing $z(\bm{s}_{i}) \sim \mathcal{N}(\mu_{i}, \sigma_{i}^2)$, where $\mu_{i} = \bm{C}_{\bm{\theta}}(\bm{s}_{i}, \mathcal{S}_{N_{i}})\mathbf{C}_{\bm{\theta}}(\mathcal{S}_{N_{i}}, \mathcal{S}_{N_{i}}^{'})^{-1}\bm{z}(\mathcal{S}_{N_{i}})$ and $\sigma_{i}^2 = C_{\bm{\theta}}(\bm{s}_{i}, \bm{s}_{i})-\bm{C}_{\bm{\theta}}(\bm{s}_{i}, \mathcal{S}_{N_{i}})\mathbf{C}_{\bm{\theta}}(\mathcal{S}_{N_{i}}, \mathcal{S}_{N_{i}}^{'})^{-1}\bm{C}_{\bm{\theta}}(\mathcal{S}_{N_{i}}, \bm{s}_{i})$.
Gibbs sampling for $\bm{z}$ is available within the generalized spatial linear model framework (\cite{Dattaetal(16a)}). Further computational insight is obtained from writing $\mathcal{N}(\bm{z}|\bm{0}, \mathbf{C}_{\bm{\theta}})$ as
\begin{align}
z(\bm{s}_{1})=\eta_{1}, \quad z(\bm{s}_{i})=a_{i,1}z(\bm{s}_{1})+a_{i,2}z(\bm{s}_{2})+\cdots+a_{i,i-1}z(\bm{s}_{i-1})+\eta_{i}\;,\;\; i=2,\ldots, n,
\end{align}
simply as $\bm{z}(\mathcal{S})=\mathbf{A}\bm{z}(\mathcal{S})+\bm{\eta}$ where $\mathbf{A}$ is $n\times n$ strictly lower-triangular with elements $a_{i,j}=0$ whenever $j\ge i$ and $\bm{\eta}\sim \mathcal{N}(\bm{0}, \mathbf{D})$ where $\mathbf{D}$ is a diagonal matrix with $d_{i,i}=Var[z(\bm{s}_{i})|\bm{z}_{<i}]$. It is obvious that $\mathbf{I}-\mathbf{A}$ is nonsigular and $\mathbf{C}=(\mathbf{I}-\mathbf{A})^{-1}\mathbf{D}(\mathbf{I}-\mathbf{A})^{-T}$.  
The neighbor selection is corresponding to introduce the sparsity into $\mathbf{A}$, i.e., $a_{i,j}\neq 0$ when $j\in N_{i}$, $a_{i,j}=0$ otherwise.  The approximated covariance matrix is obtained as $\tilde{\mathbf{C}}=(\mathbf{I}-\tilde{\mathbf{A}})^{-1}\tilde{\mathbf{D}}(\mathbf{I}-\tilde{\mathbf{A}})^{-T}$ where $\tilde{\mathbf{A}}$ is sparse approximation of $\mathbf{A}$ and the diagonal component of $\tilde{\mathbf{D}}$ is $\tilde{d}_{i,i}=Var[z_{i}|\bm{z}_{N_{i}}]$.  This can be performed in $\mathcal{O}(nM^3)$ and in parallel across rows of $\mathbf{A}$. NNGP introduces sparsity into $\tilde{\mathbf{C}}^{-1}$ not into $\tilde{\mathbf{C}}$ directly. Hence $\tilde{\mathbf{C}}$ is not necessarily sparse (unlike in covariance tapering). 
On the other hand, INLA requires $\mathcal{O}(n^{3/2})$ flops computational time and $\mathcal{O}(n\log(n))$ dynamic memory storage for a spatial case \citep{Rueetal(09)}.
\section{Inference}
\subsection{Bayesian inference in \cite{GoncalvesGamerman(18)}}
\label{sec:inf}
We follow the sampling algorithm in Section 4.2 in \cite{GoncalvesGamerman(18)}. Unknown quantities to be sampled include $\mathcal{U}, K, \lambda^{*}, \bm{z}(\mathcal{S}_{aug}), \bm{\theta}$, and we denote them by $\psi$. The joint posterior and conditional densities are
\begin{align*}
\pi(\psi|\mathcal{S})&\propto \Phi_{n}[f(\mathbf{W}_{n}, \bm{\beta}_{n})]\Phi_{m}[-f(\mathbf{W}_{m}, \bm{\beta}_{m})]\pi_{GP}(\bm{\beta}_{K}|\bm{\theta}) \nonumber \\
&\times \exp(-\lambda^{*}|\mathcal{D}|)\lambda^{*K}\frac{1}{K!}\pi(\lambda^{*})\pi(\bm{\theta}) \\
\pi(\mathcal{U}, \bm{\beta}_{m}, K|\cdot)&\propto \Phi_{m}[-f(\mathbf{W}_{m}, \bm{\beta}_{m})]\pi_{GP}(\bm{\beta}_{m}|\bm{\beta}_{n},\bm{\theta})\frac{\lambda^{*K}}{K!} \bm{1}(K\ge n), \\
\pi(\bm{\beta}_{K}|\cdot)&\propto \Phi_{n}[f(\mathbf{W}_{n}, \bm{\beta}_{n})]\Phi_{m}[-f(\mathbf{W}_{m}, \bm{\beta}_{m})]\pi_{GP}(\bm{\beta}_{K}|\bm{\theta}), \\
\pi(\lambda^{*}|\cdot)&\propto \exp\{-\lambda^{*}|\mathcal{D}| \} \lambda^{*K} \pi(\lambda^{*}), \\
\pi(\bm{\theta}|\cdot)&\propto \pi_{GP}(\bm{\beta}_{K}|\bm{\theta})\pi(\bm{\theta}), 
\end{align*}
where $m=K-n$ and $\bm{\beta}_{n}=(z(\bm{s}_{1}),\ldots,z(\bm{s}_{n}), \ldots, \beta_{p}(\bm{s}_{1}), \ldots, \beta_{p}(\bm{s}_{n}))$. Also, $\mathbf{W}_{n}=(\mathbf{I}_{n}, \mathbf{X}_{1}, \ldots, \mathbf{X}_{p})$, where $\mathbf{X}_{j}$ is an $n\times n$ diagonal matrix with the $(i, i)$-entry $\bm{X}_{j}(\bm{s}_{i})$ the $j$th covariate at location $\bm{s}_{i}$. 

\cite{GoncalvesGamerman(18)} discuss identifiability of $\lambda^{*}$. Gibbs sampling for $\lambda^{*}$ from its full conditional distribution is available when a Gamma prior is assumed, i.e., $\pi(\lambda^{*})=\mathcal{G}(\alpha, \beta)$. 
The time varying case is easily accommodated through, for example, a time dependent Gamma prior $\pi(\lambda_{t}^{*})=\mathcal{G}(\alpha_{t}, \beta_{t})$ where $\lambda_{t}^{*}$ varies independently across times. 
Another extension, which incorporates time dependence among $\lambda_{t}^{*}$s, introduces Markov structure $\lambda_{1}^{*}\sim \mathcal{G}(a_{1}, b_{1})$, $\lambda_{t}^{*}|K_{1:t-1}, \lambda_{t-1}^{*}=w^{-1}\lambda_{t-1}^{*}\zeta_{t}$ and $\zeta_{t}\sim \text{Beta}(wa_{t}, (1-w)a_{t})$, which yields tractable full conditional distributions \citep{GoncalvesGamerman(18)}. 

Updating $\bm{\theta}$ will involve space-time covariance matrix computations for which we will exploit the NNGP. 
Below, we describe implementing NNGPs for sampling $[\mathcal{U}, \bm{\beta}_{m}, K | \cdot]$ and $[\bm{\beta}_{K}|\cdot]$. 
Gibbs sampling of $\bm{\beta}_{K}|\cdot$ is based on simulating a general class of skewed normal (SN) distributions proposed by \cite{ArellanoAzzalini(06)}, see Section 4.1 in \cite{GoncalvesGamerman(18)} for details.

\noindent {\small \rule{\textwidth}{1pt}
	\textbf{Sampling $\mathcal{U}, \bm{\beta}_{m}, K|\cdot$} \\ 
	[-8pt]\rule{\textwidth}{1pt}\\[-12pt]
	\begin{enumerate}
		\item[1.] Simulate $K_{t}\sim \text{Poi}(\lambda_{t}^{*}|\mathcal{D}|)$ for $t=1,\ldots,T$. If $K_{t}=n_{t}$, make $\{\bm{u}_{j} \}_{j=1}^{m}=\emptyset$ , otherwise go to step 2. 
\item[2.] Make $j=1$ and $\bm{z}_{1:j-1}=\emptyset$
\item[3.] Make $r_{j}=1$
\item[4.] Simulate $\bm{u}_{r_{j}}\sim \text{Uniform}(\mathcal{D})$ and $\bm{\beta}(\bm{u}_{r_{j}})$ from $\pi_{GP}(\bm{\beta}(\bm{u}_{r_{j}})|\bm{\beta}_{n}, \bm{\beta}_{m}, \bm{\theta})$
\begin{itemize}
\item[(a),] Compute the column vector of distance and covariance ($\bm{C}_{\bm{\theta}}(\bm{u}_{r_{j}},\cdot)$) between $\bm{u}_{r_{j}}$ and the current locations.
\item[(b).] Compute the mean $\mu_{r_{j}}$ and variance $\sigma_{r_{j}}^2$ of the conditional GP
\begin{align}
\mu_{r_{j}}&=\mu+\bm{C}_{\bm{\theta}}(\bm{u}_{r_{j}},\cdot)\mathbf{C}_{\bm{\theta}}^{-1}(\bm{\beta}(\cdot)-\bm{\mu}) \\
\sigma_{r_{j}}^{2}&=C_{\bm{\theta}}(\bm{u}_{r_{j}}, \bm{u}_{r_{j}})-\bm{C}_{\bm{\theta}}(\bm{u}_{r_{j}},\cdot)\mathbf{C}_{\bm{\theta}}^{-1} \bm{C}_{\bm{\theta}}(\cdot, \bm{u}_{r_{j}})
\end{align}
where $\bm{\mu}$ and $\mathbf{C}_{\bm{\theta}}$ are the mean vector and covariance matrix respectively of $\pi_{GP}(\bm{\beta}(\cdot)|\bm{\theta})$.
\end{itemize}
\item[5.] Simulate $Y_{r_{j}}\sim \text{Ber}[\Phi[-f(\bm{W}(\bm{u}_{r_{j}}), \bm{\beta}(\bm{u}_{r_{j}}))]]$
\item[6.] 
\begin{itemize}
\item[(a).] If $Y_{r_{j}}=1$ and $j<K-n$, set $\bm{u}_{j}=\bm{u}_{r_{j}}$, $\mathcal{U}=\{u_{1},\ldots,u_{j}\}$, $\bm{\beta}(\bm{u}_{j})=\bm{\beta}(\bm{u}_{r_{j}})$ and update matrix $\mathbf{C}_{\bm{\theta}}^{-1}$  as follows: \hfill{$\mathcal{O}((n+j)^2)$}
\begin{align}
\mathbf{C}_{\bm{\theta}}^{-1}=\begin{pmatrix}
                                      \mathbf{C}_{\bm{\theta}}^{-1}+\mathbf{C}_{\bm{\theta}}^{-1}\bm{C}_{\bm{\theta}}(\cdot,\bm{u}_{j})(1/\sigma_{r_{j}}^2)\bm{C}_{\bm{\theta}}(\bm{u}_{j}, \cdot)\mathbf{C}_{\bm{\theta}}^{-1} & -\mathbf{C}_{\bm{\theta}}^{-1}\bm{C}_{\bm{\theta}}(\cdot, \bm{u}_{j})/\sigma_{r_{j}}^2 \\
-(1/\sigma_{r_{j}}^2)\bm{C}_{\bm{\theta}}(\bm{u}_{j},\cdot)\mathbf{C}_{\bm{\theta}}^{-1} & 1/\sigma_{r_{j}}^2
                                      \end{pmatrix}
\end{align}
Then, $j=j+1$ and go to step 3. 
\item[(b).] If $Y_{r_{j}}=1$ and $j=K-n$, set $\bm{u}_{j}=\bm{u}_{r_{j}}$, $\bm{\beta}(\bm{u}_{j})=\bm{\beta}(\bm{u}_{r_{j}})$ and go to step 8.
\item[(c).] If $Y_{r_{j}}=0$, set $r_{j}=r_{j}+1$ and go to step 4.
\end{itemize}
\item[7.] Output $\{K, \mathcal{U}=\{ \bm{u}_{j} \}_{j=1}^{m}, \bm{\beta}_{m} \}$ where $m=K-n$
	\end{enumerate}	
	\rule{\textwidth}{1pt}
\\
	\textbf{Sampling $\bm{\beta}_{K}|\cdot$} \\ 
	[-8pt]\rule{\textwidth}{1pt}\\[-12pt]
\begin{itemize}
\item[1.] Obtain $\mathbf{W}$ such that
\begin{align}
\mathbf{W}&=\begin{pmatrix}
                      \mathbf{W}_{n} & \mathbf{O}_{n\times m} \\
                      \mathbf{O}_{m\times n} & \mathbf{W}_{m}
                    \end{pmatrix}, \\
\Phi_{K}[\mathbf{W}\bm{\beta}_{K}]&=\Phi_{n}[\mathbf{W}_{n}\bm{\beta}_{n}]\Phi_{m}[-\mathbf{W}_{m}\bm{\beta}_{m}],
\end{align}
\item[2.] Sample $\bm{\beta}_{K}\sim \mathcal{SN}(\bm{\mu}, \mathbf{C}_{\bm{\theta}}, \mathbf{W})$ where $\bm{\mu}$ and $\mathbf{C}_{\bm{\theta}}$ are the mean vector and covariance matrix respectively of $\pi_{GP}(\bm{\beta}_{K}|\bm{\theta})$. 
We define $\mathbf{\Delta}^{T}=\mathbf{W}\mathbf{C}_{\bm{\theta}}$, $\bm{\gamma}=\mathbf{W}\bm{\mu}$, $\mathbf{\Gamma}=\mathbf{I}_{K}+\mathbf{W}\mathbf{C}_{\bm{\theta}}\mathbf{W}^{T}$ and $\mathbf{A}\mathbf{A}^{T}=\mathbf{\Gamma}$
\begin{itemize}
\item[(a).] Calculate $\mathbf{A}$.  
\item[(b).] Simulate a value $\bm{v}_{0}^{*}$ from $(\bm{V}_{0}^{*}|\bm{V}_{0}^{*}\in B)$ where $B=\{\bm{v}_{0}^{*}:\mathbf{A}\bm{v}_{0}^{*}>-\bm{\gamma} \}$, obtain $\bm{v}_{0}=\mathbf{A}\bm{v}_{0}^{*}$.  
\item[(c).] Simulate $\bm{\beta}_{K}$ from $(\bm{V}_{1}|\bm{V}_{0}=\bm{v})\sim \mathcal{N}(\bm{\mu}+\mathbf{\Delta}\mathbf{\Gamma}^{-1}\bm{v}, \mathbf{\Omega})$ where $\mathbf{\Omega}=\mathbf{C}_{\bm{\theta}}-\mathbf{\Delta} \mathbf{\Gamma}^{-1}\mathbf{\Delta}^{T}$.
\end{itemize}
\end{itemize}
	\rule{\textwidth}{1pt}
\\
	\textbf{Sampling $\lambda_{t}^{*}|\cdot$} \\ 
	[-8pt]\rule{\textwidth}{1pt}\\[-12pt]
\begin{itemize}
\item[1.] For $t=1,\ldots, T$, compute $a_{t}=wa_{t-1}+K_{t}$, $b_{t}=wb_{t-1}+\mu(\mathcal{S}_{t})$
\item[2.] Sample $\lambda_{t}^{*}\sim \mathcal{G}(a_{T}, b_{T})$. 
\item[3.] For $t=T-1,\ldots, 1$, sample $\lambda_{t}^{*}=w\lambda_{t+1}^{*}+L_{t}$, where $L_{t}\sim \mathcal{G}((1-w)a_{t}, b_{t}).$
\end{itemize}
	\rule{\textwidth}{1pt}
}
\subsection{NNGP implementation and computational complexity}
As for sampling $\mathcal{U}, \bm{z}(\mathcal{U}), K| \cdot$, we require $\mathcal{O}((n+j)^2)$ flops for sequentially updating $\mathbf{C}_{\bm{\theta}}^{-1}$ for $j=1,\ldots,m$. The dominant expense is $\mathcal{O}(\sum_{j=1}^{m}(n+j)^2)$. In this step, $\mathbf{C}_{\bm{\theta}}^{-1}$ is used only for simulating $z(u_{r_{j}})$. NNGP does not require calculation of $\mathbf{C}_{\bm{\theta}}^{-1}$, $z(u_{r_{j}})$ is generated from $\mathcal{N}(\mu_{r_{j}}, \sigma_{r_{j}}^{2})$, where $\mu_{r_{j}}=\mu+\bm{C}_{\bm{\theta}}(\bm{u}_{r_{j}},\mathcal{S}_{N_{r_{j}}})\mathbf{C}_{\bm{\theta}}(\mathcal{S}_{N_{r_{j}}}, \mathcal{S}_{N_{r_{j}}}^{'})^{-1}(\bm{z}(\mathcal{S}_{N_{r_{j}}})-\bm{\mu}_{N_{r_{j}}})$ and $\sigma_{r_{j}}^{2}=C_{\bm{\theta}}(\bm{u}_{r_{j}}, \bm{u}_{r_{j}})-\bm{C}_{\bm{\theta}}(\bm{u}_{r_{j}},\mathcal{S}_{N_{r_{j}}})\mathbf{C}_{\bm{\theta}}(\mathcal{S}_{N_{r_{j}}}, \mathcal{S}_{N_{r_{j}}}^{'})^{-1} \bm{C}_{\bm{\theta}}(\mathcal{S}_{N_{r_{j}}}, \bm{u}_{r_{j}})$, 
and $N_{r_{j}}$ is the set of neighbors of $\bm{u}_{r_{j}}$. Computing $\mathbf{C}_{\bm{\theta}}(\mathcal{S}_{N_{r_{j}}}, \mathcal{S}_{N_{r_{j}}}^{'})^{-1}$ require $\mathcal{O}(M^3)$ for each $j$, so the dominant cost with NNGP is $\mathcal{O}(mM^3)$ \citep{Finleyetal(17)}.

As for sampling $\bm{z}(\mathcal{S}_{aug})|\cdot$, the exact approach requires calculating $\mathbf{\Gamma}^{-1}$ and simulating $\bm{v}\sim \mathcal{N}(\bm{0}, \mathbf{\Gamma})$ on $B$ whose computational cost is $\mathcal{O}(K^3)$ flops. This is practically unfeasible even with moderate $n$ because $K$ is always greater than $n$. Calculating $\mathbf{\Gamma}^{-1}$ using NNGP requires $\mathcal{O}(KM^3)$ flops and the same computational cost for $\mathbf{A}$.  In practice, for sampling $\bm{v}_{0}$ under the restriction $B=\{\bm{v}_{0}^{*}:\mathbf{A}\bm{v}_{0}^{*}>-\bm{\gamma} \}$ we prefer sequential updating of $v_{0,i}$ for $i=1,\ldots, K$. Using NNGP, we sequentially update $v_{0,i}\sim \mathcal{N}(\mu_{0,i}, \sigma_{0,i}^2)$, where 
\begin{align*}
\mu_{0,i}&=\bm{\Gamma}_{\bm{\theta}}(\bm{s}_{i},\mathcal{S}_{N_{i}})\mathbf{\Gamma}_{\bm{\theta}}(\mathcal{S}_{N_{i}}, \mathcal{S}_{N_{i}}^{'})^{-1}\bm{v}_{0}(\mathcal{S}_{N_{i}}), \quad \bm{s}_{i}\in \mathcal{S}_{aug} \\
\sigma_{0,i}^{2}&=\Gamma_{\bm{\theta}}(\bm{s}_{i}, \bm{s}_{i})-\bm{\Gamma}_{\bm{\theta}}(\bm{s}_{i},\mathcal{S}_{N_{i}})\mathbf{\Gamma}_{\bm{\theta}}(\mathcal{S}_{N_{i}}, \mathcal{S}_{N_{i}}^{'})^{-1} \bm{\Gamma}_{\bm{\theta}}(\mathcal{S}_{N_{i}}, \bm{s}_{i})\;,
\end{align*}
then accept $v_{0,i}$ when $v_{0,i}>-\gamma_{i}$ for $i=1,\ldots, K$. Furthermore, we also need samples from $\bm{v}_{1}\sim \mathcal{N}(\bm{\mu}+\Delta \mathbf{\Gamma}^{-1}\bm{v}_{0}, \mathbf{\Omega})$. In this step, no further NNGP approximation is required for $\mathbf{\Omega}$. 
We sequentially update $\bm{v}_{1}\sim \mathcal{N}(\mu_{1,i}, \sigma_{1,i}^2)$, where 
\begin{align*}
\mu_{1,i}&=\mu_{i}+\mathbf{\Delta}(\bm{s}_{i}, \mathcal{S}_{N_{i}})\mathbf{\Gamma}_{\bm{\theta}}(\mathcal{S}_{N_{i}}, \mathcal{S}_{N_{i}}^{'})^{-1} \bm{v}_{0}(\mathcal{S}_{N_{i}}) \\
\sigma_{1,i}^{2}&=C_{\bm{\theta}}(\bm{s}_{i}, \bm{s}_{i})-\mathbf{\Delta}(\bm{s}_{i}, \mathcal{S}_{N_{i}})\mathbf{\Gamma}_{\bm{\theta}}(\mathcal{S}_{N_{i}}, \mathcal{S}_{N_{i}}^{'})^{-1}\mathbf{\Delta}(\mathcal{S}_{N_{i}}, \bm{s}_{i}) 
\end{align*}

Finally, the computational cost incurred when NNGP is used for exGCP is dominated by inversion of $K$ matrices, each of order $M\times M$, in sampling $\bm{z}(\mathcal{S}_{aug})|\cdot$. This step can be parallelized across $K$ processors. So, the computational cost is further reduced to $\mathcal{O}(KM^{3}/J)$, where $J$ is the number of available cores/threads.   

\section{Simulation Examples}
In this section, we investigate recovering the intensity for spatial exGCP and space-time exGCP with NNGP approximations. All the simulations for our methodology are coded in Ox \citep{Doornik(07)} and run on Intel(R) Xeon(R) Processor X5675 (3.07GHz) with 12 Gbytes of memory.
\subsection{Example 1: spatial Gaussian Cox processes}
We investigate recovering the intensity surface using an NNGP model. We assume $\mathcal{D}=[0,10]\times [0,10]$, $\lambda^{*}=20$, $\bm{W}(\bm{s})=1$ and $\bm{\beta}(\bm{s})=z(\bm{s})$, and define the model as   
\begin{align}
f(\bm{W}(\bm{s}), \bm{\beta}(\bm{s}))=z(\bm{s}), \quad \bm{z}\sim \mathcal{N}(\bm{0}, \mathbf{C}_{\bm{\theta}})\;,
\end{align}
where $\mathbf{C}_{\bm{\theta}}=[\sigma^2\exp(-\phi \|\bm{s}_{i}-\bm{s}_{j} \|)]_{i,j=1,\ldots,n}$ and $\bm{\theta}=(\sigma^2, \phi)$. We set $\sigma^2=1$ and $\phi=2$ and fix these parameter values for inference. First, we simulate $\mathcal{S}_{homo}$ from a homogeneous Poisson process on $\mathcal{D}$ with intensity $\lambda^{*}|\mathcal{D}|$, i.e., $\mathcal{S}_{homo} \sim HPP(\lambda^{*}|\mathcal{D}|)$, and $\bm{z}(\mathcal{S}_{homo})\sim \mathcal{N}(\bm{0}, \mathbf{C}_{\bm{\theta}}(\mathcal{S}_{homo}, \mathcal{S}_{homo}'))$. Then, we retain locations $\bm{s}_{i}\in \mathcal{S}_{homo}$ with probability $\Phi[f(\bm{W}(\bm{s}_{i}), \bm{\beta}(\bm{s}_{i}))]$, denote $\mathcal{S}$ for the set of retained points, i.e., the realization from the point process with intensity $\lambda(\bm{s}_{i})=\lambda^{*}\Phi[f(\bm{W}(\bm{s}_{i}), \bm{\beta}(\bm{s}_{i}))]$ on $\bm{s}_{i}\in \mathcal{D}$. The number of points in $\mathcal{S}$ is $n=1086$. 

\begin{figure}[ht]
  \begin{center}
   \includegraphics[width=16cm]{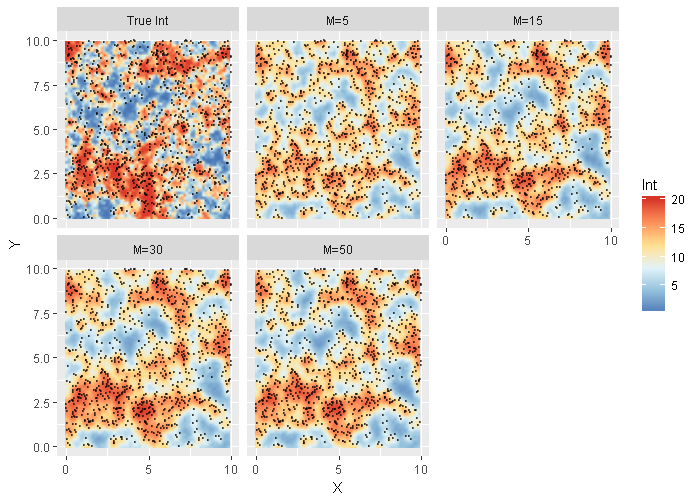}
  \end{center}
  \caption{The intensity surface $\lambda^{*}\Phi[f(\bm{W}(\cdot), \bm{\beta}(\cdot))]$: true (top left), $M=5$ (top middle), $M=15$ (top right), $M=30$ (bottom left) and $M=50$ (bottom middle). The black dots are simulated $\mathcal{S}$}
  \label{fig:Sim1}
\end{figure}

Turning to Bayesian inference, we run MCMC by devising a joint Gibbs sampler for the latent Gaussian variables. We monitored the chains for convergence and, in particular, calculated the inefficiency factor (reciprocal of effective sample size). Each draw of the sampler constitutes one realization from the exact multivariate posterior distribution. Hence, convergence is expectedly rapid and this is corroborated from calculating the inefficiecy factor (inverse of effective sample size). In fact, we monitored the 500 iterations and found that they yielded almost 500 independent realizations of the multivariate latent variables from the joint posterior distribution. We feel that this is adequate for calculating the marginal means and variances for each latent variable. Indeed, our sampler does not require such a long burn-in period, also consistent with findings by \cite{GoncalvesGamerman(18)}, and a burn-in of 100 initial samples was deemed sufficient.

We fix hyperparameters $\bm{\theta}$ and $\lambda^{*}$ at true values because the likelihood does not have much information for these parameters \citep{GoncalvesGamerman(18)}.
As for NNGP, we consider four cases, $M=5$, $M=15$, $M=30$ and $M=50$, to investigate the accuracy of the approximation. Figure \ref{fig:Sim1} plots $\mathcal{S}$ and the true and estimated intensity surface $\lambda^{*}\Phi[f(\bm{W}(\cdot), \bm{\beta}(\cdot))]$ for $M=5$, $M=15$, $M=30$ and $M=50$. 
Computational time for each case is 21.5 min ($M=5$), 25.7 min ($M=15$), 28.2 min ($M=30$) and 45.9 min ($M=50$).
The estimated intensity surfaces are smoother than the true intensity surface, but these surfaces are almost indistinguishable from each other. For example, the maximum difference is $\text{max}_{i}|\hat{\lambda}(\bm{s}_{i}|M=30)-\hat{\lambda}(\bm{s}_{i}|M=50)|=3.638$ where $\hat{\lambda}(\bm{s}_{i}|M)$ is the posterior mean of the intensity at $\bm{s}_{i}$ with $M$ neighbors. Clearly, $M=30$ is more than sufficient for substantive inference.

 %
 %

\subsection{Example 2: spatial-time Gaussian Cox processes}
\begin{figure}[ht]
  \begin{center}
   \includegraphics[width=16cm]{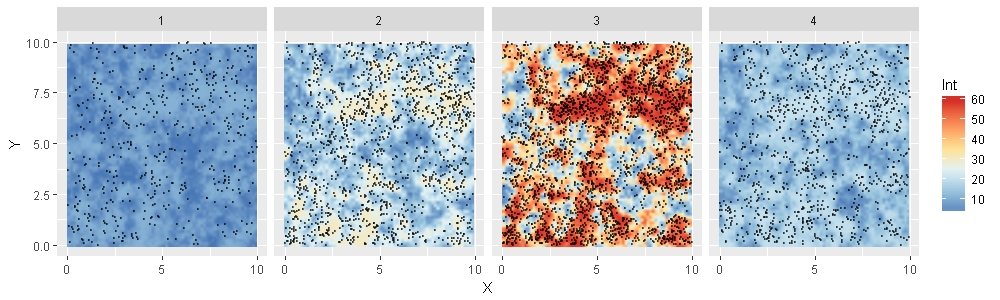}
  \end{center}
  \caption{The intensity surface $\lambda^{*}\Phi[f(\bm{W}(\cdot), \bm{\beta}(\cdot))]$. The black dots are simulated $\mathcal{S}$}
  \label{fig:Sim3int}
\end{figure}

Next, we investigate recovering the intensity surface for a space-time case with time varying $\lambda_{t}^{*}$.  
Again, we assume $\mathcal{D}=[0,10]\times [0,10]$ and $T=4$. 
The model is defined as
\begin{align*}
\lambda_{t}&=\lambda_{t}^{*}\Phi[f(\bm{W}_{t}(\bm{s}), \bm{\beta}_{t}(\bm{s}))], \quad f(\bm{W}_{t}(\bm{s}), \bm{\beta}_{t}(\bm{s}))=z_{t}(\bm{s}) \\
\bm{z}_{t}&=\bm{z}_{t-1}+\bm{\eta}_{t}, \quad \bm{\eta}_{t}\sim \mathcal{N}(\bm{0}, \mathbf{C}_{\bm{\theta}_{t}}), \quad \text{for} \quad t=2, 3, 4 \\
\bm{z}_{1}&=\bm{\eta}_{1}, \quad \bm{\eta}_{1}\sim \mathcal{N}(\bm{0}, \mathbf{C}_{\bm{\theta}_{1}})
\end{align*}
where $\bm{\theta}_{1}=(\sigma_{1}^2, \phi_{1})$ and $\bm{\theta}_{2}=\bm{\theta}_{3}=\bm{\theta}_{4}=(\sigma^2, \phi)$. 
We set $(\sigma_{1}^2,\phi_{1}, \sigma^2,\phi)=(1,2, 0.3, 3)$ and fix these parameter values for inference. We assume that the time varying $\lambda_{t}^{*}$ are $(\lambda_{1}^{*}, \lambda_{2}^{*}, \lambda_{3}^{*}, \lambda_{4}^{*})=(10,30,60,20)$. 
The number of simulated points are $(n_{1}, n_{2}, n_{3}, n_{4})=(513, 1540, 3207, 1075)$. Although the pattern itself is similar across time, the number of points fluctuate sharply. 
Figure \ref{fig:Sim3int} exhibits a simulated space-time point pattern $\mathcal{S}$ and true intensity surface for $t=1,\ldots, 4$. 

\begin{figure}[ht]
  \begin{center}
   \includegraphics[width=16cm]{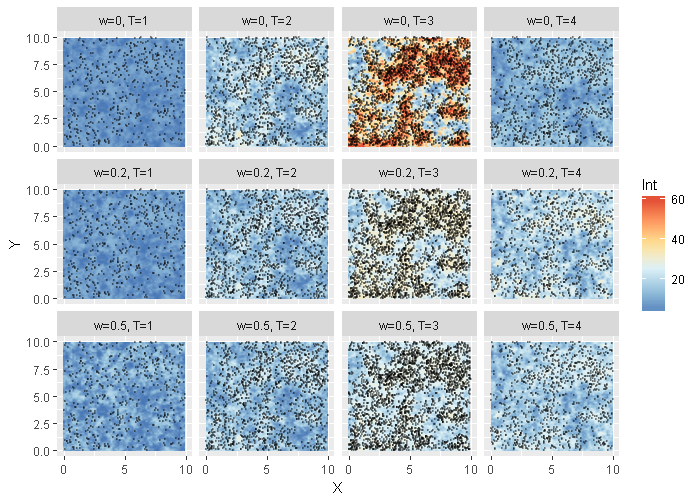}
  \end{center}
  \caption{The plot of the estimated intensity surface $\lambda^{*}\Phi[f(\bm{W}(\cdot), \bm{\beta}(\cdot))]$ for $t=1,\ldots, 4$. $w=0$ (top), $w=0.2$ (middle), $w=0.5$ (bottom)}
  \label{fig:Sim3est}
\end{figure}

Again, we run MCMC, discard the first 100 samples as burn-in and retain the next 500 samples as posterior samples. We set $M=30$ for the number of neighbors. 
We assume the time varying prior specifications for $\bm{\lambda}^{*}$ introduced in Section \ref{sec:inf}, set $a_{0}=100, b_{0}=10$. We note that $\lambda_{t}^{*}$ is sensitive to the choice of $w$: larger values indicate stronger persistence. We produce the estimated intensity surface under different $w$ settings. Figure~\ref{fig:Sim3est} shows the estimated intensity surface $\lambda^{*}\Phi[f(\bm{W}(\cdot), \bm{\beta}(\cdot))]$ for $w=0$, $w=0.2$ and $w=0.5$, where $w=0$ corresponds to an independent prior $\lambda_{t}^{*}$. The true intensity surface is well recovered in this case. When $w=0.2$ and $w=0.5$, the scale of the intensity surface is smoothed across time, i.e., degraded for large points, upgraded for small points.

\subsection{Example 3: spatial-time Gaussian Cox processes: real data settings}
\label{sec:sim3}
\begin{figure}[ht]
  \begin{center}
   \includegraphics[width=16cm]{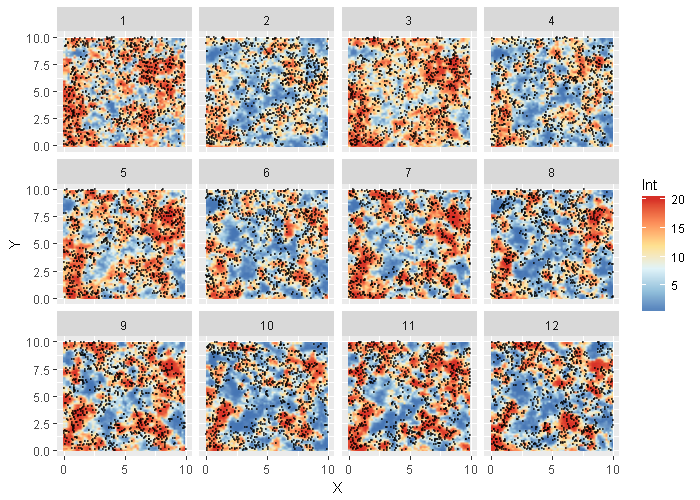}
  \end{center}
  \caption{The true intensity surface $\lambda^{*}\Phi[f(\bm{W}_{t}(\cdot), \bm{\beta}_{t}(\cdot))]$ for $t=1,\ldots, T$. The black dots are simulated $\mathcal{S}$}
  \label{fig:Sim2int}
\end{figure}

Finally, we investigate another space-time setting similar to real data in Section \ref{sec:real}. 
We assume $\mathcal{D}=[0,10]\times [0,10]$ and $T=12$. 
\begin{align*}
\lambda_{t}&=\lambda_{t}^{*}\Phi[f(\bm{W}_{t}(\bm{s}), \bm{\beta}_{t}(\bm{s}))], \quad f(\bm{W}_{t}(\bm{s}), \bm{\beta}_{t}(\bm{s}))=z_{t}(\bm{s}) \\
\bm{z}_{t}&=\bm{z}_{t-1}+\bm{\eta}_{t}, \quad \bm{\eta}_{t}\sim \mathcal{N}(\bm{0}, \mathbf{C}_{\bm{\theta}}), \quad \text{for} \quad t=2, \ldots, 12 \\
\bm{z}_{1}&=\bm{\eta}_{1}, \quad \bm{\eta}_{1}\sim \mathcal{N}(\bm{0}, \mathbf{C}_{\bm{\theta}_{1}})
\end{align*}
Considering real data in Section \ref{sec:real}, time invariant $\lambda^{*}$ is a reasonable assumption for simulating datasets. 
We set $(\sigma_{1}^2,\phi_{1}, \sigma^2,\phi)=(1,2,0.3, 3)$ and fix these parameter values for inference.
The total number of points is $11,581$, $n_{t}$ range from 933 to 1023. 
Figure \ref{fig:Sim2int} is the true intensity surface $\lambda^{*}\Phi[f(\bm{W}_{t}(\cdot), \bm{\beta}_{t}(\cdot))]$ on $\mathcal{S}_{t}$ for $t=1,\ldots, T$.

\begin{figure}[ht]
  \begin{center}
   \includegraphics[width=16cm]{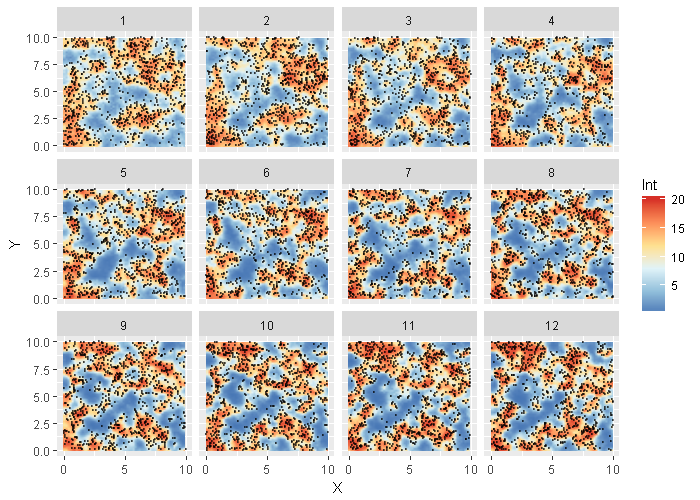}
  \end{center}
  \caption{The posterior mean of the intensity surface $\lambda^{*}\Phi[f(\bm{W}_{t}(\cdot), \bm{\beta}_{t}(\cdot))]$ for $t=1,\ldots, T$. The black dots are simulated $\mathcal{S}$}
  \label{fig:Sim2est}
\end{figure}

As for Bayesian inference, we run MCMC, discarding the first 100 samples as a burn-in, preserving the subsequent 500 samples as posterior samples. We consider $M=30$ for the number of neighbors. 
We assume time varying prior specification for $\bm{\lambda}^{*}$, set $a_{0}=200$, $b_{0}=10$ and $w=0.5$.
Figure~\ref{fig:Sim2est} depicts the posterior mean intensity surface $\lambda^{*}\Phi[f(\bm{W}_{t}(\cdot), \bm{\beta}_{t}(\cdot))]$ for $t=1,\ldots,T$.
The estimated surface is smoother than the true surface but captures the behavior of the true surface well, as also seen in previous examples. 

\begin{figure}[ht]
  \begin{center}
   \includegraphics[width=16cm]{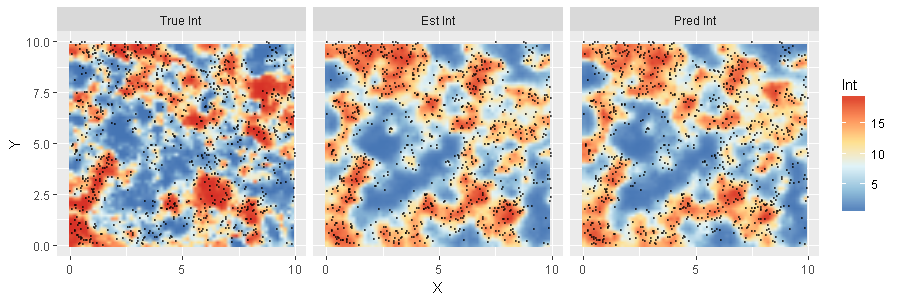}
  \end{center}
  \caption{The true and posterior mean of the estimated and predictive intensity surface $\lambda^{*}\Phi[f(\bm{W}_{t}(\cdot), \bm{\beta}_{t}(\cdot))]$ at $t=12$: true (left), estimated (middle) and predictive (right),}
  \label{fig:Sim2pred}
\end{figure}

Finally, we demonstrate the prediction results for $t=12$. 
Predictive surface is recovered with predictive distribution $\lambda_{12}^{*}$. Our time series structure implies $\bm{z}_{12, pred}(\cdot)=\bm{z}_{11}(\cdot)$, i.e., posterior predictive mean of spatial random field at $T=12$ is posterior mean of spatial random field  at $t=11$.   
Figure \ref{fig:Sim2pred} is the true, estimated and predictive intensity surface at $t=12$.
The true and estimated intensity surface at $t=12$ are the same in Figure \ref{fig:Sim2int} and \ref{fig:Sim2est}, respectively.
The estimated and predictive intensity surfaces show similar patterns including their scales.

\section{Real Data Application: Crime Event Data in San Francisco}
\label{sec:real}
\begin{figure}[ht]
  \begin{center}
   \includegraphics[width=15cm]{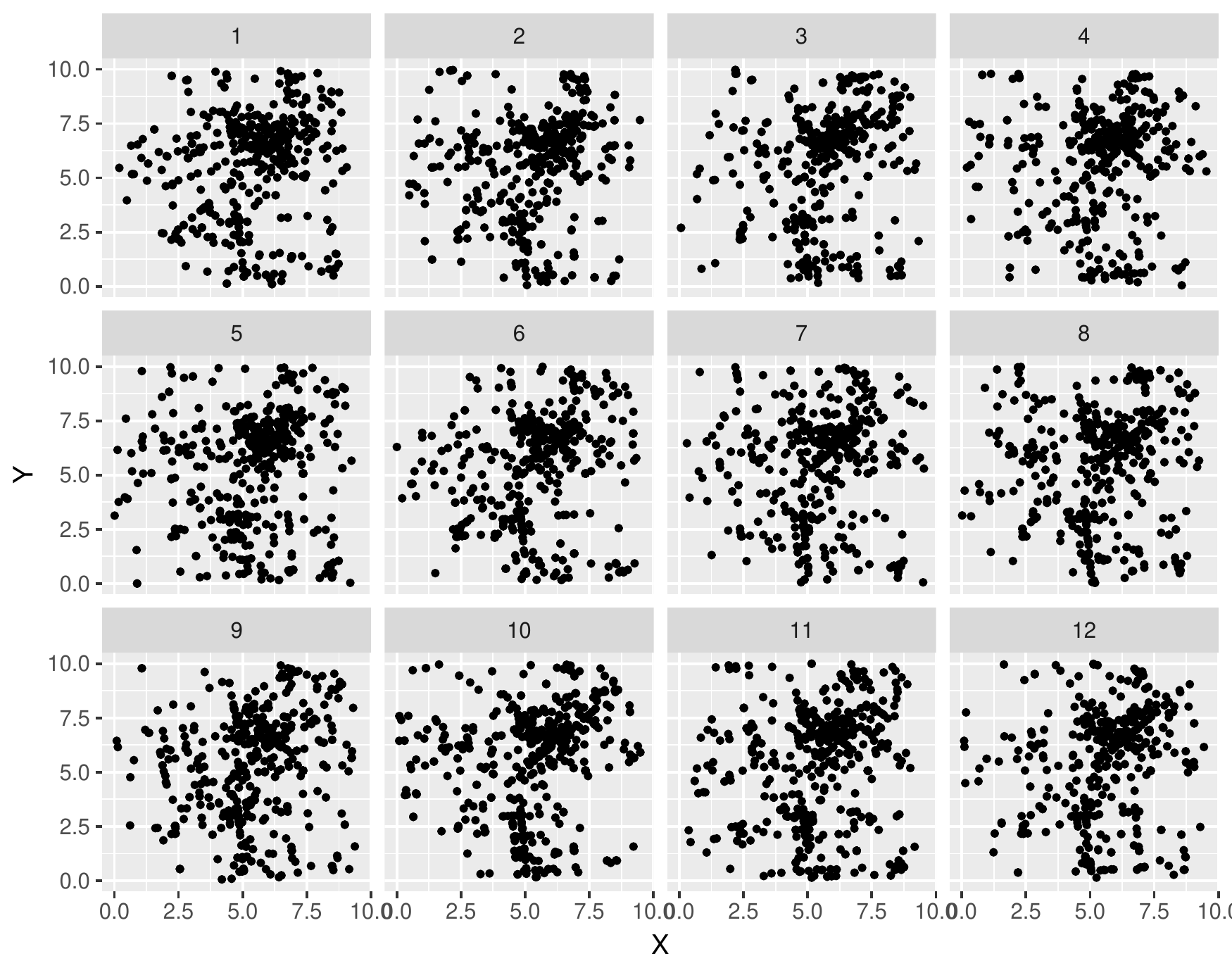}
  \end{center}
  \caption{The plot of transformed Assault events $\mathcal{S}$ in $\mathcal{D}$.}
  \label{fig:Realplot}
\end{figure}

Our dataset consists of crime events in the city of San Francisco (SF) in 2012. We focus on Assault events in the rectangular region $\mathcal{D}=[-122.45, -122.39]\times [37.75, 37.800]$ which is surrounding the Tenderloin district, where lots of crime events are observed. 
We transform longitude and latitude information into easting and northing information, and project them onto $\mathcal{D}=[0,10]\times [0,10]$.  Figure \ref{fig:Realplot} is the plot of transformed Assault events in 2012. The data contains $6,174$ points, $n_{t}$ range from 481 to 582. Unfortunately, no covariate information is available. We take twelve months ($T=12$) as the time index and investigate monthly crime event patterns. Across the months, point patterns exhibit similar behavior, especially concentration around $[5,7.5]\times [5,7.5]$. This kind of large clustering of points requires large $K$ relative to the number of observed points ($n$), i.e., $n\ll m$.

\begin{figure}[ht]
  \begin{center}
   \includegraphics[width=16cm]{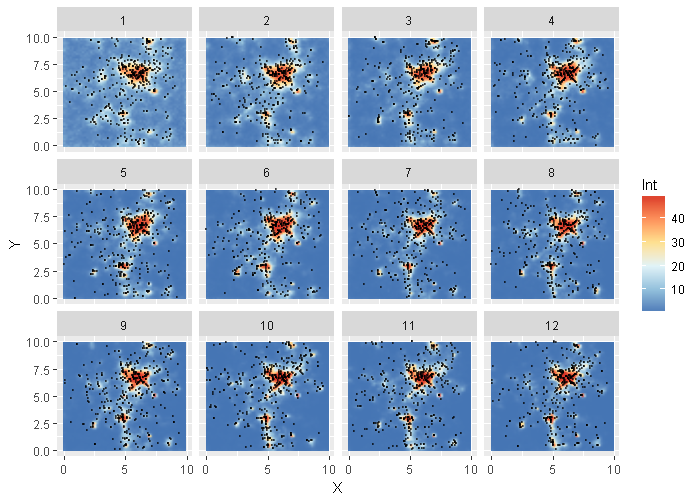}
  \end{center}
  \caption{The posterior mean of the intensity surface $\lambda^{*}\Phi[f(\bm{W}_{t}(\cdot), \bm{\beta}_{t}(\cdot))]$. The block dots are $\mathcal{S}$.}
  \label{fig:Realest}
\end{figure}

Our model specification is the space-time Gaussian Cox process, investigated with a simulation example in Section \ref{sec:sim3}, which is defined as 
\begin{align*}
\lambda_{t}&=\lambda_{t}^{*}\Phi[f(\bm{W}_{t}(\bm{s}), \bm{\beta}_{t}(\bm{s}))], \quad f(\bm{W}_{t}(\bm{s}), \bm{\beta}_{t}(\bm{s}))=z_{t}(\bm{s}) \\
\bm{z}_{t}&=\bm{z}_{t-1}+\bm{\eta}_{t}, \quad \bm{\eta}_{t}\sim \mathcal{N}(\bm{0}, \mathbf{C}_{\bm{\theta}}), \quad \text{for} \quad t=2, \ldots, 12 \\
\bm{z}_{1}&=\bm{\eta}_{1}, \quad \bm{\eta}_{1}\sim \mathcal{N}(\bm{0}, \mathbf{C}_{\bm{\theta}_{1}})\; .
\end{align*}
We set $\bm{\theta}_{1}=(\sigma_{1}^2, \phi_{1})=(1,2)$ $\bm{\theta}=(\sigma^2, \phi)=(0.3, 3)$, which are selected through pre-processed runs of the algorithm. 
We also introduce time varying $\lambda_{t}^{*}$, and set $a_{0}=500$, $b_{0}=10$ and $w=0.5$. 
Since the prior expectation of the number of points is $\lambda_{1}^{*}|\mathcal{D}|=5,000$, the computational cost without any approximation is about $\mathcal{O}(TK^3)$ flops, where $K\approx 5,000$ for each MCMC iteration. 
This will be unfeasible within modest computing environments. Again, we take $M=30$ nearest neighbors as $N_{i}$ for $i=1,\ldots,K$. 
Our inference is again based on $500$ posterior samples retained after discarding the first $100$ samples as pre-convergence burn-in. Figure \ref{fig:Realest} is the posterior mean intensity surface $\lambda^{*}\Phi[f(\bm{W}(\cdot), \bm{\beta}(\cdot))]$. As demonstrated in simulation studies, the posterior mean explains the clustering property of the crime event patterns while capturing local behavior.

\begin{figure}[ht]
  \begin{center}
   \includegraphics[width=16cm]{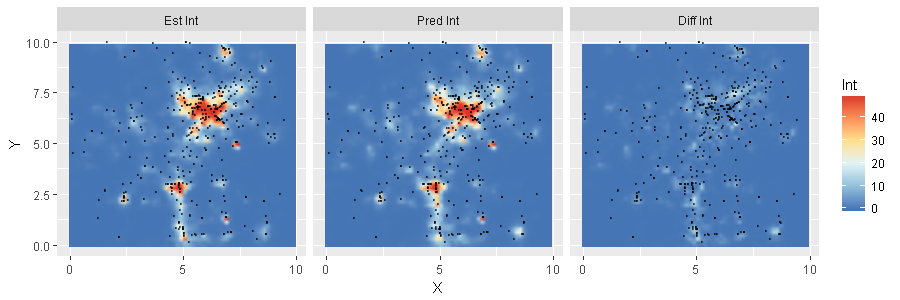}
  \end{center}
  \caption{The posterior mean of the estimated (left) and predictive (middle) intensity surface $\lambda^{*}\Phi[f(\bm{W}_{t}(\cdot), \bm{\beta}_{t}(\cdot))]$ at $t=12$ and their absolute difference (right). The block dots are $\mathcal{S}_{12}$.}
  \label{fig:Realpred}
\end{figure}

Finally, we check the posterior predictive intensity surface at $t=12$. The predictive intensity surface is $\hat{\lambda}_{12}^{*}\Phi[\hat{\bm{z}}_{12}(\cdot)]$, where $\hat{\lambda}_{12}^{*}$ is simulated from posterior predictive distribution defined in Section 4.1, and $\hat{\bm{z}}_{12}(\cdot)$ is the posterior mean of $\bm{z}_{11}(\cdot)$. Figure \ref{fig:Realpred} is the posterior mean of the estimated and predictive intensity surface and their absolute difference. The estimated intensity surface has the same intensity at $t=12$ in Figure \ref{fig:Realest}. The maximum value of the absolute difference is $19.95$. The estimated and predictive intensity surfaces show similar patterns including their scales except for some local variations.

\section{Discussion}
This paper proposes a specific computationally efficient implementation for space-time Gaussian Cox processes proposed by \cite{GoncalvesGamerman(18)} using the NNGP as described in \cite{Dattaetal(16a)}. We demonstrate that our method captures the intensity surfaces well, while keeping moderate computational costs for relatively large point patterns. Inference is performed via MCMC, in particular the Gibbs sampler. We implement our algorithm for crime event data in San Francisco which has a larger number and cluster of points than examples in \cite{GoncalvesGamerman(18)}. The number of neighbors for the NNGP is specified by the user and, as shown in our simulations, fairly small numbers of neighbors usually suffice to capture the substantive features of the surface. The recovered intensity surface is robust to the choice of the number of neighbors through simulation studies. 

Future work will implement our algorithm for space and {\it continuous} time with nonseparable space-time covariance functions as detailed in \cite{Dattaetal(16b)}. 
Without any approximation of covariance functions, sampling the Gaussian process is implausible for nonseparable space-time covariance function. Our approach is promising for such settings.  We will also evaluate biases caused by approximating and comparing practical computational times of exGCP with other existing approaches in a comprehensive way. 
\section*{Acknowledgement}
The computational results are obtained by using Ox version 7.1 \citep{Doornik(07)}.
The authors thank Fl\'{a}vio B. Gon\c{c}alves for providing the Ox code in \cite{GoncalvesGamerman(18)}.

\bibliographystyle{chicago}
\bibliography{SP}
\end{document}